\documentclass[preprint]{emulateapj}
\usepackage{epsfig} 
\usepackage{epstopdf}
\usepackage{graphicx}
\usepackage{amsmath} 
\usepackage{hyperref}
\usepackage{tabularx}
\usepackage{aas_macros}

\usepackage{breakcites}

\newcommand{\HII}{\mbox{H II}}

\newcolumntype{R}{>{\centering\arraybackslash}X}

\begin{document}

\title{The Discovery of Raman Scattering in \HII\ Regions}

\author{Michael A. Dopita\altaffilmark{1}, David C. Nicholls\altaffilmark{1}, Ralph S. Sutherland\altaffilmark{1},  Lisa J. Kewley\altaffilmark{1} \& Brent A. Groves\altaffilmark{1}}
\email{Michael.Dopita@anu.edu.au}

\altaffiltext{1}{Research School of Astronomy and Astrophysics, Australian National University, Canberra, ACT 2611, Australia.}

\begin{abstract}
We report here on the discovery of faint extended wings of H$\alpha$ observed out to an apparent velocity of $\sim 7600$\,km\,s$^{-1}$ in the Orion Nebula (M42) and in five \HII\ regions in the Large and the Small Magellanic Clouds. We show that, these wings are caused by Raman scattering of both the O\,I and Si\,II resonance lines and stellar continuum UV photons with H\,I followed by radiative decay to the H\,I $n=2$ level. The broad wings also seen in H$\beta$ and in H$\gamma$ result from Raman scattering of the UV continuum in the H\,I $n=4$ and $n=5$ levels respectively.The Raman scattering fluorescence is correlated with the intensity of the narrow permitted lines of O\,I and Si\,II. In the case of Si\,II, this is explained by radiative pumping  of the same 1023.7\AA\ resonance line involved in the Raman scattering by the Ly$\beta$ radiation field. The subsequent radiative cascade produces enhanced  Si\,II $\lambda \lambda 5978.9, 6347.1$ and 6371.4\AA\ permitted transitions. Finally we show that in O\,I, radiative pumping of the 1025.76\AA\ resonance line by the Lyman series radiation field is also the cause of the enhancement in the permitted lines of this species lying near H$\alpha$ in wavelength, but here the process is a little more complex. We argue that all these processes are active in the zone of the \HII\ region near the ionisation front.
\end{abstract}

 
\section{Introduction}
In the observations of \HII\ regions, broad H$\alpha$ lines with widths in excess of  100\,km\,s$^{-1}$ have been reported in a number of papers. For example, \citet{Castaneda90} reported a broad feature with FWHM$\sim 500$\,km\,s$^{-1}$ in the H$\alpha$ and [N\,II] emission lines of the giant extragalactic \HII\ region NGC\,5471, while for an extragalactic \HII\ region in NGC\,2363, \citet{Roy92} discovered an even wider feature with FWHM$\sim2400$\,km\,s$^{-1}$. This is seen in H$\alpha$,  H$\beta$ and in the [O\,III] lines. In both these cases, the broad features are present in the forbidden as well as the recombination lines, showing that we are dealing with excitation of high-velocity plasma such as stellar winds or supernova remnants. 

Certainly, fast bulk motions powered by mechanical energy input from the exciting stars has been identified by \citet{Rozas06} in bright, isolated \HII\ regions in spiral galaxies. However, the velocities involved here are $\lesssim 100$\,km\,s$^{-1}$. For objects displaying broader features such as NGC\,2363, \citet{Binette09} developed a plausible model based upon photo-evaporating dense clouds embedded in a fast stellar wind. Here turbulent mixing and entrainment of the photo-evaporated material into the stellar wind flow was shown to reproduce the observed line profiles in both the forbidden and the recombination lines.

Up to the present, broad H$\alpha$ features with widths in excess of  1000\,km\,s$^{-1}$ have not been reported in `normal' \HII\ regions excited by one or a few central stars. In this paper, we provide very high dynamic range spectrophotometry with good resolution ($R=7000$) using the Wide-Field Spectrograph \citep[WiFeS,][]{Dopita07,Dopita10} for six high-surface brightness, young, but otherwise normal \HII\ regions around single O-stars or small clusters. Five of these are located in the Large Magellanic Cloud (LMC) or the Small Magellanic Cloud (SMC), and we also have data on a Galactic \HII\ region, the Orion Nebula (M42). For all six objects observed, we find broad Lorentzian tails underlying the H$\alpha$ line extending out to an apparent radial velocity of at least $\sim7600$\,km\,s$^{-1}$. Here we demonstrate that these wings are not the result of bulk motions, but arise as a consequence of Raman scattering fluorescence of O\,I and Si\,II resonance lines and stellar UV continuum which excite H\,I to a virtual level near the Ly$\beta$ transition. This is the first time this process has been observed in \HII\ regions.

\section{Observations}
The spectroscopic data were obtained the Wide-Field Spectrograph. This instrument, an image-slicing double-beam integral-field spectrograph, is mounted at the Nasmyth focus of the Australian National University 2.3m telescope located at the Siding Spring Observatory, Australia.  It provides 25$\times$38 spaxels each $1\times$\,1\arcsec\ in angular size. We used the $R=7000$ gratings, in two separate setups which allows us to construct continuous spectra with adequate overlap of coverage between gratings. Our data cover the full wavelength range $3300-8950$\AA\, at a velocity resolution of $\Delta v =45\,$km\,s$^{-1}$.

The observational data were obtained during two runs covering the nights of October 21-23, 2014, and November 10-15, 2015. In the case of M42, we observed a bright but moderately featureless region centered at RA = 05hr 35m 12.3s, Dec = -05:22:24 (J2000), with the long axis of the WiFeS aperture oriented E--W. 
All the remaining objects fitted entirely within the WiFeS field of view, allowing us to extract their global spectra.

In order to properly correct for saturation of the CCD in the cores of the strong lines, stepped exposure times were used (e.g. 10s, 100s and 800s in Orion and 20s, 60s and 900s in N88A). Absolute photometric calibration of the data cubes was made using the STIS spectrophotometric standard stars HD\,009051, HD\,031128 and HD\,200654\footnote{Fluxes available at: \newline {\url{www.mso.anu.edu.au/~bessell/FTP/Bohlin2013/GO12813.html}}}.  In addition, the B-type stars HIP\,18926, HIP\,111085, and HIP\,106768 were used to provide improved telluric calibration. Separate corrections for the OH and H$_2$O telluric absorption features were made.  Arc and bias frames were also taken regularly, and internal continuum lamp flat fields and twilight sky flats were taken in order to provide sensitivity corrections in both the spectral and spatial directions.

The data were reduced using the {\tt PyWiFeS v0.7.3} pipeline written for the instrument \citep{Childress14a, Childress14b}. In brief, this produces a data cube which has been wavelength calibrated, sensitivity corrected (including telluric corrections), photometrically calibrated, and from which the cosmic ray events have been removed. Because the  {\tt PyWiFeS} pipeline uses an optical model of the spectrograph to provide the wavelength calibration, the wavelength solution is good across the whole field, and does not rely on any interpolation of the data, since each spaxel is assigned a precise wavelength and spatial coordinate. 

From each data cube, we extracted integral spectra of the \HII\ region in the case of the LMC and SMC objects based on the observed spatial extent of the nebula in H$\alpha$. In the case of the Orion nebula, we extracted the spectrum from most of the data cube, avoiding regions of suspected emission line saturation or regions known to contain the shock-excited Herbig-Haro knot, HH\,202. A complete description of the spectrophotometry extracted from these spectra and the extraction technique will be given in a future paper.

\section{The Broad wings on H$\alpha$}
\begin{figure}[htb!]
\begin{centering}
\includegraphics[scale=0.48]{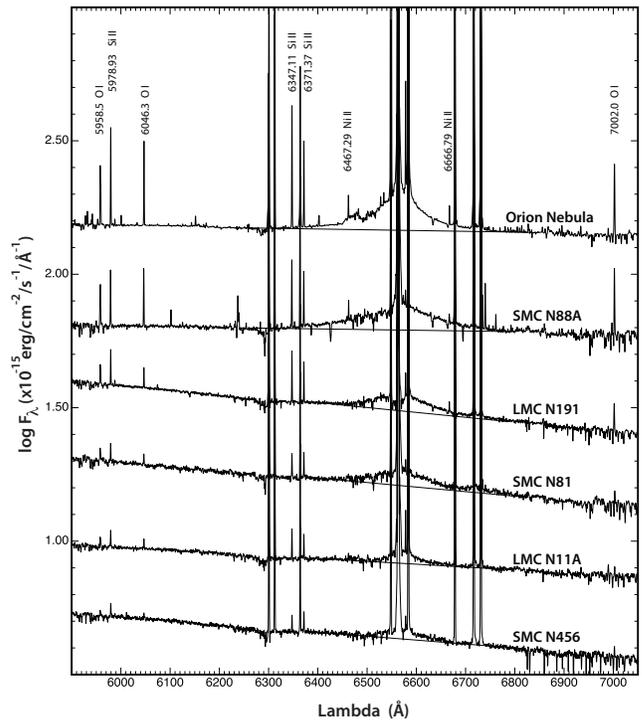}
\end{centering}
\caption{The observation of broad wings around H$\alpha$ in six bright \HII\ regions. The data are presented on a log scale, and have been offset in flux in order to separate the spectra and to order them by strength of the broad wings. In the objects with the strongest features, these wings can be traced at least as far as 6730\AA\ which, if interpreted as a velocity would correspond to 7600\,km\,s$^{-1}$. Note that in Orion, there is a secondary broad feature at about 6475\AA\, and that in LMC N191 there is a pronounced dip in the broad component near the line center, which may be stellar absorption. The continuum is a mixture of nebular continuum and flux from the exciting stars which causes the tilt in the baseline.} \label{fig1}
\end{figure}

\begin{figure}[htb!]
\begin{centering}
\includegraphics[scale=0.43]{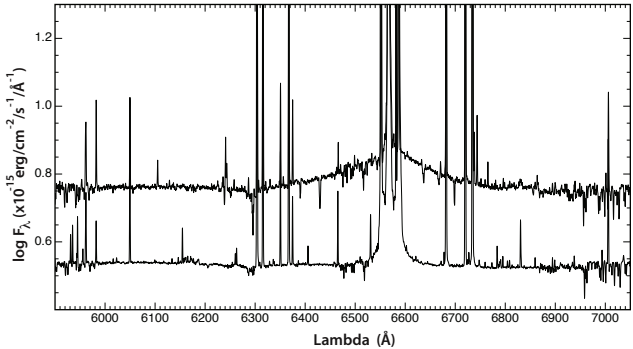}
\end{centering}
\caption{A comparison of SMC N88A with the very bright low-excitation planetary nebula IC\,418. Note the absence of broad wings of H$\alpha$ in the IC\,418 spectrum} \label{fig2}
\end{figure}

In each of the six \HII\ regions we observed, we identified low-level broad wings in H$\alpha$ which were not present in other emission lines of similar intensity. This is shown in Figure \ref{fig1} where the spectra have been reduced to rest wavelength, and ordered in decreasing intensity of the broad component. As we demonstrate below, these wings can be well fit by a Lorentzian function. In the case of the Orion nebula, these wings are also visible (but not remarked upon) in the MUSE spectrum presented by \citet{Weilbacher15}. In Orion, we measure the peak flux to be only $1.3\times10^{-4}$ of the main H$\alpha$ line. The noise in the background nebular continuum is $<10^{-6}$ of the peak H$\alpha$ intensity. In the case of Orion, we also detect broad wings on H$\beta$, and in the case of SMC N88A we see broad wings in H$\beta$ and, faintly, in H$\gamma$ as well. We can eliminate the possibility that these wings are of instrumental origin through the following arguments:
\begin{itemize}
\item{The design of the WiFeS instrument is such as to keep ghosts, scattering or grating artefacts to a very low level. The slicer itself fully fills the field with less than 5\,$\mu$m spaces between the 1.75\,mm slices. In addition, the light from each slice passes through a slice mask to eliminate scattered light. Furthermore the transmissive VPH gratings provide very high spectral purity. Finally, the camera employs optimised anti-reflection coatings, and is designed to eliminate ghost images at the CCD.}
\item{The wings are still visible in the short exposure images, where the H$\alpha$ line is not saturated in its core, eliminating the possibility that  CCD saturation is generating an artefact.}
\item{The wings are not visible in other objects with similar spectra and surface brightness. In Figure \ref{fig2}, we present a comparison of SMC N88A with the low-excitation planetary nebula IC\,418. Despite the strength of the [N\,II] and H$\alpha$ lines in this object, there is no evidence of the extended wings seen in SMC N88A.}
\end{itemize}

\section{Raman Scattering}
Raman scattering is the process whereby resonance line photons generated by  an ion of one atom excite a second ion of a different atom (usually Hydrogen) to a virtual state, immediately followed by radiative decay to another state in the same ion. This process is strongly favoured when the excited virtual state lies very close to a quantised state in the second ion, since the virtual state density increases rapidly as a quantised state is approached. { The effect of this is that the Raman scattering cross section closely approximates to a Lorenzian function centred on the resonant (Rayleigh) wavelengths of the scattering atom. This was given for the specific specific example of Hydrogen by  \citet{Nussbaumer89}.}

The process was first observed (but not identified) in novae by \citet{Joy45} and \citet{Thackeray74} and in symbiotic stars by \citet{Allen76}. In these objects broad lines are seen at 6830\AA\ an 7090\AA. These lines were first identified by \citet{Schmid89} as being the result of Raman scattering of the O\,VI resonance doublet $\lambda\lambda1032,1038$ by excitation to virtual states of hydrogen lying close to the Ly$\beta$ transition, followed by radiative decay to the excited 2s$^2$S state. Since then, O\,VI Raman scattering has been described in many papers, and due to other resonances such as the He\,II $\lambda 1025$ line, see \citet{Sekeras15} and references therein. Furthermore, Raman scattered He\,II features were identified in Planetary Nebulae (PNe) by \citet{Pequignot97} and these have since been identified in other PNe \citep{Groves02,Lee06,Kang09}. O\,VI Raman scattering was also discovered in the Planetary Nebula NGC\,6302 by \citet{Groves02}. 

A detailed description of the Raman scattering process by emission lines is to be found in \citet{Nussbaumer89}, where a number of other candidate ions which may initiate Raman scattering are also given, including Raman scattering of the O\,I resonance line. Since a UV line is scattered into an optical frequency photon, this process multiplies the original difference in wavelength between the  Raman scattered transition producing broad H$\alpha$ wings and the Ly$\beta\ \lambda1025$ level by approximately the ratio of the wavelengths of  H$\alpha$ to Ly$\beta$, or 6.4. It also multiplies intrinsic line widths and profiles by a similar factor.

The theoretical solution to the problem of the formation of broad wings by Raman scattering in Symbiotic stars is given by \citet{Jung04} who showed the importance of Raman Scattering of the stellar UV continuum in the formation of broad wings. This theory was later applied to PNe  by \citet{Lee09}, and to Active Galaxies by \citet{Chang15}. In this theory, the width of the stellar UV Raman-scattered wings depends on the column density of neutral hydrogen and the shape of the far wings can be approximated by $F_{\lambda} \propto \left(\lambda-\lambda_{\mathrm Ly_{\beta}}\right)^{-2}$. Raman scattering of the stellar UV continuum may also account for the broad wings of H$\beta$ and H$\gamma$, which result from excitation of H\,I into virtual levels near the $n=4$ and $n=5$ states, respectively.\vspace{0.4cm}

\subsection{O\,I Raman scattering}
In the case of the \HII\ regions observed here, the formation of the broad H$\alpha$ feature is a combination of stellar UV Raman scattering { (which for a flat EUV continuum spectrum has a Lorentzian profile), and Raman scattering by O\,I (for which the shape of the Raman Scattered profile primarily reflects the shape of the  O\,I resonance line)}:
\begin{equation}
\mathrm{O\,I}~~ 2p^3 3d~^3\mathrm{D}^0_{1,2,3}\stackrel{\rm \lambda1025.76}{\rightarrow } 2p^4~{^3\mathrm{P}}_2\;  \label{1}
\end{equation}
for this transition, the (vacuum) wavelength lies very close to that of the Ly$\beta$ line; $ \lambda1025.72$. The { slight difference between the two ensures that Raman scattered feature due to this process is centered (in air) at 6565.38\AA.}

We approximate the shape of the combined Raman scattered feature by a Lorentzian function, since this gives an accurate description to the shape of the far wings of H$\alpha$, and is well behaved closer to the core. { However, since the stellar EUV Raman scattering approximated to a Lorenz function, and the Raman scattered O\,I line is that of an optically-thick resonance line profile, a better approximation to the summed (O\,I + Stellar UV) Raman scattering profile is more like a Voigt function.} Unfortunately this could not be constrained near the line centre due to the presence of the strong H$\alpha$ and [N\,II] emission lines, which are a factor $\sim10^4$ times stronger than the Raman feature.  The parameters of our Lorentzian fits are given in Table \ref{Table1}. Note that, as predicted,  the peak wavelengths are displaced red-ward of the peak of the H$\alpha$, although the observed shift is somewhat greater than the prediction, at an average rest wavelength of 6568.4\AA. This is probably in part the result of the residual telluric absorption, which preferentially affects the blue wing of the broad feature. However \citet{Jung04} also predict the H$\alpha$ wing profiles to have a higher asymmetry on the red side than the blue side. Such an effect would also help to produce the peak shift we see in our simple line fitting. Finally, velocity shifts can also be produced by relative motion between the H\,I-- and the O\,I--emitting gas. However, given that these two ions are strongly coupled to each other by charge exchange, such shifts are likely to be very small.

The variation in the width of the Lorentz fit reflects the variation in the H\,I column density between different objects according to the \citet{Jung04} theory. The ratio of the peak of the broad feature to the narrow line, given in the last column of Table \ref{Table1}, is remarkably constant -- lying between $0.9\times10^{-4}$ and  $1.8\times10^{-4}$. {  However, the Raman Scattering profile arises from the combination of the local
stellar EUV continuum flux (which is HII region geometry dependent), and from the density of H I atoms (which is both ionisation and pressure dependent). The H$\alpha$ emission line itself arises simply from recombination throughout the ionised nebula. Thus our observation that the peak flux ratio is constant within a factor of three is not very constraining on the physics.}

\begin{table*}
  \centering
  \caption{Measured Raman scattering parameters in \HII\ Regions}\label{Table1}
  {\scriptsize
   \begin{tabular}{llcccc}
\hline
   {\bf O I + Stellar UV Raman line:} &   &   &     &    &    \\
    Object: &  Measured & Rest  & Peak Flux  & Lorentz &Peak Flux  Ratio  \\
               &  $\lambda$(\AA) &$\lambda$(\AA) & (erg/cm$^2$/s$^{-1}$\AA$^{-1}) $  & FWHM (\AA) & (wrt. H$_\alpha$) \\
\hline
    Orion (M42)  &  6566.7$\pm$0.3 & 6566.4$\pm$0.3  & 9.30$\pm$0.2 E-14 & 91$\pm$8    & 1.1E-4 \\
    LMC N11A   & 6577.5$\pm$1.5  &  6570.6$\pm$1.6 & 2.90$\pm$0.3 E-15 & 71$\pm$6    & 1.0E-4  \\
    LMC N191A & 6573.2$\pm$2.5  & 6567.4$\pm$2.6  & 2.25$\pm$0.5 E-15 & 109$\pm$5 &1.4E-4  \\
    SMC N81A   & 6572.8$\pm$2.5  & 6569.0$\pm$2.6  & 1.45$\pm$0.2 E-15 & 98$\pm$10 & 0.9E-4 \\
    SMC N88A   & 6571.6$\pm$2.5  & 6567.6$\pm$2.6  & 3.25$\pm$0.5 E-13 & 137$\pm$20 &  1.8E-4   \\
    SMC N456   & 6574.2$\pm$2.5  & 6569.7$\pm$2.6  & 1.65$\pm$0.7 E-15 & 87$\pm$5   & 1.5E-4  \\
\hline
    {\bf Si II Raman line:} &   &   &    &    &  \\
     Object: &  Measured & Rest  & Peak Flux  & Gaussian & Peak Flux  Ratio\\
              &  $\lambda$(\AA) &$\lambda$(\AA) & (erg/cm$^2$/s$^{-1}$\AA$^{-1}) $  & FWHM (\AA) & (wrt. H$_\alpha$)  \\
\hline 
   Orion (M42)  &  6472.4$\pm$1.5    & 6472.1$\pm$1.6    & 1.7$\pm$0.2 E-14 & 27.0$\pm$2.5 & 2.1E-5  \\
    LMC N11A &  6486.0$\pm$3        &   6479.0$\pm$3.0     & 1.1$\pm$0.2 E-15 & 19$\pm$6        &   3.8E-5  \\
   SMC N88A &  6475.5$\pm$3        &   6471.5$\pm$3.0    & 6.6$\pm$0.2 E-16 & 36$\pm$6        &   3.6E-5  \\
    \hline
\end{tabular}%
}
\end{table*}
\begin{figure*}[htb!]
\begin{centering}
\includegraphics[scale=0.55]{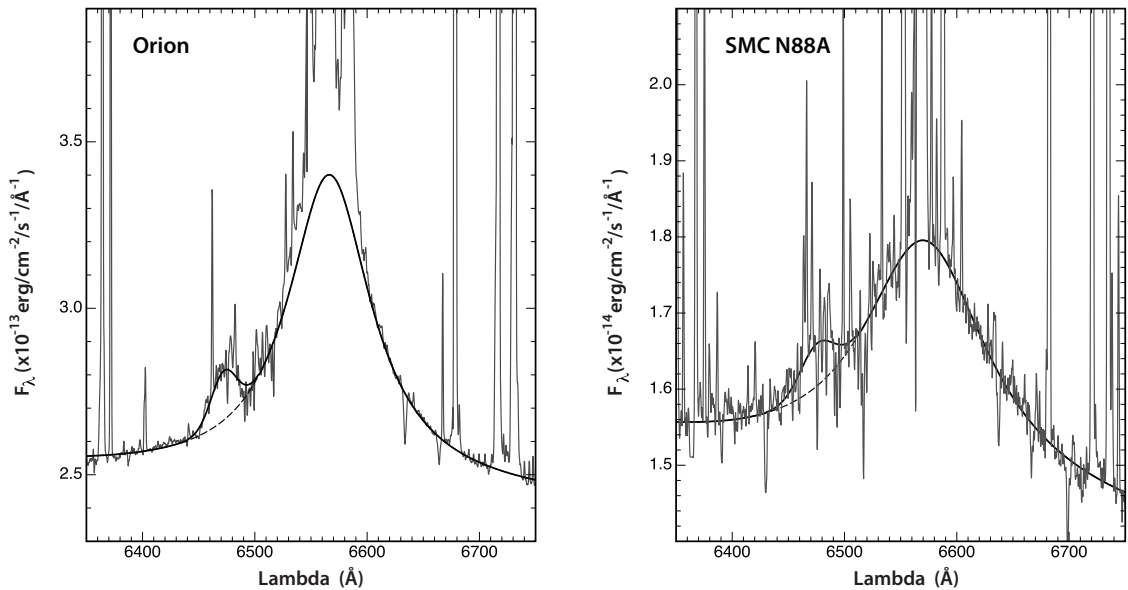}
\end{centering}
\caption{The fit to the Raman scattering profiles of O\,I and Si\,II in the Orion Nebula (M42) and in SMC N88A. The O\,I Raman profile is best fit with a Lorentzian, the broad wings of which result from Raman scattering of the stellar UV continuum (see text), while the Si\,II bump is better fit with a simple Gaussian.} \label{fig3}
\end{figure*}

\subsection{Si\,II Raman scattering}
We have also identified a second Raman scattering which most likely gives rise to the secondary broad feature -- evident in the Orion nebula at about 6475\AA, and also detected in SMC N88A and LMC N11A. Here Raman scattering is caused by the Si\,II line:
\begin{equation}
\mathrm{Si\,II}~~ {3s^2}5s~^2\mathrm{S_{1/2}} \stackrel{\rm \lambda1023.70}{\rightarrow }{3s^2}3p~{^2\mathrm{P}}^{o}_{3/2} \label{2}
\end{equation}
In this case we expect the Raman scattered `H$\alpha$' line for this process to be centred (in air) at 6480.1\AA. This feature is  detectable only in three objects, M42 , LMC N11A, and SMC N88A, and is best fit to the observations by a Gaussian; see Table \ref{Table1}. The width of the Raman features suggest an intrinsic line width in the Si\,II 1023.7\AA\ line of 0.46 -- 0.87\AA. Again, the ratio of the peak flux in the Si\,II  feature to the peak H$\alpha$ flux is almost constant, in the range $2.1-3.8\times10^{-5}$. The mean rest wavelength of the peak of this feature (6474.2\AA) is somewhat smaller than the theoretical prediction, but again it is affected by the residual Telluric H$_2$O absorption, which in this case lies in the red wing of the feature.  In Figure \ref{fig3} we show the quality of the fits to the O\,I and Si\,II Raman scattering profiles for the two objects showing the strongest features, Orion and SMC N88A. The region of Telluric  H$_2$O absorption referred to above lies in the approximate wavelength range 6470-6570\AA.

\section{Radiative Pumping of O\,I and Si\,II Lines}
A striking feature of Figure \ref{fig1} is the correlation between the Raman scattered wing intensity, and the intensities of the O\,I $\lambda\lambda$ 5958.6, 6046.3 and 7002.0\AA\ lines, and the Si\,II $\lambda\lambda$ 5978.9, 6347.1 and 6371.4\AA\ permitted lines. Both of these correlations are related to the fact that these species co-exist in a region with a strong  Ly$\beta$ radiation field, and therefore can be radiatively pumped by this radiation field to their excited states, as we will now demonstrate.

\subsection{The Si\,II Permitted lines}
For the Si\,II lines, the observed enhancement in the permitted lines is directly caused by Ly$\beta$ radiative pumping to the excited state,
\begin{equation}
3s^2 3p~{^2\mathrm{P}}^{o}_{3/2} \stackrel{\rm Ly\beta}{\rightarrow } 3s^2  5s~^2\mathrm{S}_{1/2}\,
\end{equation}
which then gives rise to the radiative cascade:
\begin{eqnarray}
3s^2  5s~^2\mathrm{S}_{1/2}  \stackrel{\lambda 5978.9}{\rightarrow } 3s^2  4p~^2\mathrm{P}^o_{3/2} \\
3s^2  4p~^2\mathrm{P}^0  \stackrel{\lambda\lambda 6347.1,~6371.4}{\rightarrow } 3s^2  4s~^2\mathrm{S} \\
3s^2  4s~^2\mathrm{S}  \stackrel{\lambda\lambda 1526.7,~1533.4}{\rightarrow } 3s^2  3p~^2\mathrm{P}^{o} ~.
\end{eqnarray}
Since the transition which leads to the Raman scattering is the same that is involved in radiative pumping, it is therefore not surprising that the strength of the Raman-scattered wings is correlated to the strength of the lines involved in the cascade from the radiatively pumped upper level.

\subsection{The O\,I Permitted lines}
The mechanism enhancing the intensity of the O\,I lines is a little more complicated than for the Si\,II lines, but again is ultimately the result of strong pumping by the local Lyman series radiation field. In this case, radiative pumping of the O\,I transition,
\begin{equation}
2s^2 2p^4~{^3\mathrm{P}}_2 \stackrel{\rm Ly\beta}{\rightarrow } 2s^2 2p^3 3d~^3\mathrm{D}^0_{1,2,3}\, \label{OI}
\end{equation}
is followed either by the radiative cascade
\begin{eqnarray}
2s^2 2p^3 3d~^3\mathrm{D}^0_{1,2,3} \stackrel{\rm \lambda11290}{\rightarrow } 2s^2 2p^3 3p~^3\mathrm{P}\\
2s^2 2p^3 3p~^3\mathrm{P} \stackrel{\rm \lambda8448.6} {\rightarrow } 2s^2 2p^3 3s~^3\mathrm{S}^0\\
2s^2 2p^3 3s~^3\mathrm{S}^0 \stackrel{\rm \lambda1302.2} {\rightarrow } 2s^2 2p^4~{^3\mathrm{P}} ~.
\end{eqnarray}
{ or else by direct decay back to the ground state. When the optical depth of O\,I transition is high, the population in the excited state is effectively Case B, and the only effective radiative decay path is through the radiative cascade given above. The build-up of the population of the $3p~^3\mathrm{P}$ state due to radiative pumping gives rise to a very strong enhancement in the strength of the fluorescent $\lambda8448.6$ line. A detailed description of this process was developed by \citet{Kastner95}. In the case of M42, our observed flux in this line exceeds 2\% of H$\beta$, confirming that the $2
p^4~^3\mathrm{P} - 3d~^3\mathrm{D}^0$ transition is optically thick.

To explain the enhancement in the $\lambda\lambda$ 7002.0, 5958.6, and 6046.3\AA transitions is somewhat more complex, since these cannot be directly radiatively fed from the $^3\mathrm{D}^0$ state.  However, if the radiative pumping by Ly$\beta$ into this state is high, then its population will be much higher than in LTE. Such rapid radiative pumping is made possible by the very close coincidence in wavelength between Ly$\beta~\lambda1025.722$ and the O I resonance line at $\lambda1025.762$. The difference in wavelength between these is 0.04\AA\, which is less than the thermal plus turbulent width of the H I line corresponding to 0.075\AA\ in this part of the nebula \citep{Dopita72}. This raises the possibility of collisional excitation into still more highly excited states, such as the $4d~^3\mathrm{D}^0$, $5d~^3\mathrm{D}^0$ and $6s~^3\mathrm{S}^0$ levels. Each of these can then radiatively decay to the $3p~^3\mathrm{P}$ state, giving rise to the observed transitions at $\lambda\lambda$ 7002.0, 5958.6, and 6046.3\AA, respectively.  

\section{Conclusions}
Using very high signal to noise and extremely high dynamic range integral field data obtained for six normal \HII\  regions we have established, for the first time, the existence of Raman scattering of O\,I, Si\,II and of the stellar UV continuum with H\,I in these regions. Furthermore, we have both found, and explained a correlation between this Raman scattering and the enhancement of permitted lines by radiative pumping of O\,I and Si\,II into excited states by the Lyman series emission and the stellar EUV radiation field. 

The transition zone near the ionisation fronts of these \HII\ regions can provide the appropriate conditions to support these Raman scattering processes. First, both O\,I and Si\,II co-exist at high fractional ionisation with the H\,I in this zone. Second, the temperature is lower here and the atom density higher, which increases the recombination rate. This is also an essential condition to support the high photon density in the  Ly$\beta$ radiation field which is required for radiative pumping, since it only takes a few scatterings to degrade a Ly$\beta$ photon into a Ly$\alpha$ plus an H$\alpha$ photon. Thirdly, the column density of  H\,I is high, increasing the probability of Raman scattering. A high column density also supports the production of the broad Raman scattering wings around H$\alpha$ as has been demonstrated theoretically by  \citet{Jung04}.\vspace{0.2cm}

\section*{Acknowledgements}
Dopita, Kewley and Nicholls acknowledge the support of the Australian Research Council (ARC) through Discovery project DP130103925. Groves acknowledges the support of the Australian Research Council as the recipient of a Future Fellowship (FT140101202). This research has made use of \textsc{SAOImage DS9} \citep{Joye03}, developed by Smithsonian Astrophysical Observatory.

\bibliographystyle{apj}

\end{document}